\documentclass[]{spie}
\usepackage[]{graphicx}

\title{Measurement of non-Gaussian shot noise: influence of the environment} 
\author{B. Reulet\supit{1,2}, L. Spietz\supit{1}, C.M. Wilson\supit{1},
J. Senzier\supit{1} and D.E. Prober\supit{1}
\skiplinehalf
\supit{1}Departments of Applied Physics and Physics\\
 Yale University,  New Haven CT 06520-8284, USA\\
\supit{2}Laboratoire de Physique des Solides, UMR8502, b\^at 510\\
Universit\'e Paris-Sud, 91405 Orsay, France\\
}

\begin{document}
\maketitle

\begin{abstract}

We present the first measurements of the third moment of the voltage fluctuations in a conductor. 
This technique can provide new and complementary information on the electronic transport in conducting systems. The measurement was performed on non-superconducting tunnel junctions as a function of voltage bias, for various temperatures and bandwidths up to 1GHz. The data demonstrate the significant effect of the electromagnetic environment of the sample.
\end{abstract}

\vspace{3mm}
\section{INTRODUCTION}
\vspace{2mm}

Transport studies provide a powerful tool for investigating electronic properties of a conductor.  The $I(V)$ characteristic (or the differential resistance $R_{diff}=dV/dI$) contains partial information on the mechanisms responsible for conduction.  A much more complete description of transport in the steady state, and further information on the conduction mechanisms, is given by the probability distribution of the current $P$, which describes both the dc current $I(V)$ and the fluctuations. Indeed, even with a fixed voltage $V$ applied, $I(t)$ fluctuates, due to  
the discreteness of the charge carriers, the probabilistic character of scattering and the fluctuations of the population of energy levels at finite temperature $T$ \cite{BuBlan}.

The current fluctuations are characterized by the moments of the probability distribution $P$ of order two and higher. Experimentally, the average over $P$ is obtained by time averaging. Thus, the average current is the dc current $I=\left<I(t)\right>$, where $\left<.\right>$ denotes time average. The second moment (the variance) of $P$, $\left<\delta I^2\right>$, measures the amplitude of the current fluctuations, with $\delta I(t)=I(t)-I$. The third moment $\left<\delta I^3\right>$ (the skewness) measures the asymmetry of the fluctuations
\footnote{$\left< \delta I^3\right>=\left<\left<I^3\right>\right>$, the third cumulant of refs.~\citenum{Lesovik,LevitovMath,LevRez,roche,Yuval,Nagaev,Kindermann,Kindermann2,Shelankov,Buttiker,Chtche}}.
Gaussian noise $P(I)\propto{\rm exp}(-\alpha \delta I^2)$ is symmetric, so it has no third moment.
The existence of the third moment is related to the breaking of time reversal symmetry by the dc current; at zero bias, $I=0$ and positive and negative current fluctuations are equivalent, so $\left<\delta I^3\right>=0$.
An intense theoretical effort has emerged recently to calculate the third and higher moments of $P$ in various systems \cite{Lesovik,LevitovMath,LevRez,roche,Yuval,Nagaev,Kindermann,Kindermann2,Shelankov,Buttiker,Chtche}. However, until now only the second moment has been measured in the many systems studied. 
Interest in the third and higher moments has occurred, first, because its character is predicted to differ significantly from that of the second moment; in particular, the third moment is insensitive to the sample 's own Johnson noise for the voltage bias case, yet is more sensitive, in a very subtle manner, to noise and loading by the environment.  Second, measurements of the higher moments may provide a new tool for studying conduction physics, complementary to the second moment.  It is the first set of issues that we address.

In this article we summarize the first measurements of the third moment of the voltage fluctuations across a conductor,  $\left<\delta V^3\right>$, where $\delta V(t)=V(t)-V$ represents the voltage fluctuations around the dc voltage $V$ (some of the data presented here have been published in Ref.~\citenum{nous}). Below we relate this to $\left<\delta I^3\right>$.  Our experimental setup is such that the sample is current biased at dc and low frequency but the electromagnetic environment has an impedance $\sim50\;\Omega$ within the detection bandwidth, 10 MHz to 1.2 GHz. We have investigated tunnel junctions because they are predicted to be the simplest system having asymmetric current fluctuations. However, any kind of good conductor can be studied with the techniques we have developed. We studied two different samples, at liquid helium, liquid nitrogen and room temperatures. Our results are in agreement with a recent theory that considers the strong effect of the electromagnetic environment of the sample \cite{Kindermann2}. Moreover, we show that certain of these environmental effects can be dramatically reduced by signal propagation delays from the sample to the amplifier. This can guide future experiments on more exotic samples.

\vspace{3mm}
\section{THEORETICAL OVERVIEW}
\vspace{2mm}
\subsection{PERFECT VOLTAGE BIAS}

We present first the case of voltage bias. 
The scattering theory applied to transport (the so-called Landauer picture) has been very succesful in predicting many properties of phase-coherent systems coupled to reservoirs.  We thus employ that framework. The two contacts to the sample are reservoirs at temperature $T$. The sample is described by a set of $M$ transmission channels characterized by their transmission probability $T_n$, which we take to be energy independent. Except for the quantum point contact \cite{Urbina}, only statistical properties of $\left\{T_n\right\}$ can be accessed. The case of a diffusive wire is special in that it is well described by random matrix theory, and the full probability distribution of the transmission coefficients is known \cite{Dorokhov}. The tunnel junction of area $\cal A$ with a uniform barrier of tranparency $t$ can be modeled by $M\sim2k_F^2\cal A$ channels, each of which has $T_n=t\ll1$. Until now only the first two moments ${\cal T}_p=M^{-1}\sum_n (T_n)^p$ of the probability distribution of $\left\{T_n\right\}$ have been accessed in measurements: the conductance $G=M{\cal T}_1e^2/h$, which involves the average ${\cal T}_1$ of the $\left\{T_n\right\}$, and the noise power density $S_{I^2}$ (in A$^2/$Hz) :
\begin{equation}
S_{I^2}=e\eta GV{\rm coth}(eV/2k_BT)+2k_BT(1-\eta)G
\label{eq_S2}
\end{equation}
where $\eta=1-({\cal T}_2/{\cal T}_1)$ is the Fano factor, which involves the variance of $\left\{T_n\right\}$. For an opaque tunnel junction, all $T_n\ll1$ and $\eta=1$; for a diffusive wire $\eta=1/3$. At equilibrium, $V=0$ and Eq.(\ref{eq_S2}) reduces to the Johnson noise $S_{I^2}=2k_BTG$, whereas for $eV\gg k_BT$ (the so called shot noise regime), Eq.(\ref{eq_S2}) leads to $S_{I^2}=e\eta I$
\footnote{One often finds $S_{I^2}(V=0)=4k_BTG$ and  $S_{I^2}(V\gg k_BT/e)=2e\eta I$ in the literature; the factor 2 corresponds to the contributions of positive and negative frequencies.}.

For $S_{I^3}$, we extend the case of a voltage biased single channel \cite{LevRez} to obtain for a multichannel conductor:  
\begin{equation}
S_{I^3}(V,T)=e\eta Gk_BT\left(6\hat tf({\cal U})+(1-2\hat t){\cal U}\right)
\label{eq_S3}
\end{equation}
(in A$^2/$Hz$^3$) with ${\cal U}=eV/k_BT$ and $\hat t=({\cal T}_2-{\cal T}_3)/({\cal T}_1-{\cal T}_2)$. We find that the function $f({\cal U})$ is simply related to $S_{I^2}$ by $dS_{I^2}/dV=e\eta Gf({\cal U})$. The  tunnel junction with uniform barrier of transparency $t$ corresponds to $\hat t=t$, the diffusive wire to $\hat t=2/5$ \cite{Yuval,Nagaev}. We can rewrite Eq. (\ref{eq_S3}) as:

\begin{equation}
S_{I^3}(V,T)=6\hat tk_BT\frac{dS_{I^2}}{dV}+(1-2\hat t)\eta e^2GV
\label{eq_S3bis}
\end{equation}

The first term of Eq. (\ref{eq_S3bis}) is temperature dependent, whereas the second term is simply proportionnal to the current and completely temperature independent.

We consider some limits of Eq. (\ref{eq_S3bis}). 
For $eV\ll k_BT$, $S_{I^3}=e^2\eta GV$. There is no dependence on temperature.  For an opaque tunnel junction, $S_{I^3}=e^2I$.  This result is also obtained for classical, uncorrelated quantized charges of magnitude $e$ passing across a barrier.  The correlations due to the Fermion reservoirs are seen only in the factor $\eta$, which is less than one only when some $T_n$ are finite. The magnitude of $S_{I^3}$ for $eV\sim2k_BT$ is of order $e\eta Gk_BT$, i.e., scales with the temperature. For $eV\gg k_BT$, Eq. (\ref{eq_S3bis}) gives $S_{I^3}=e\eta G[k_BT+eV(1-2\hat t)]$.
It can be negative when $\hat t>1/2$, as for the single quantum channel at large voltage when $t>1/2$.  Thus, if more than half the electrons incident from the left reservoir are transmitted, $S_{I^3}$ is negative.  Still, $S_{I^3}$ is always positive for $eV\ll k_BT$, and is positive at all voltages if all $T_n\ll1$ (in which case $\eta\sim1$ and $\hat t\ll1$).

It is noticeable that the \emph{shape} of the $S_{I^2}(V,T)$ curve does not depend solely on $\eta$, so a careful calibration is necessary to  obtain $\eta$. In contrast,
 $\hat t$ can be experimentally extracted from the ratio $r$ of the slopes at high voltage and low voltage; $r=1-2\hat t$. Thus $\hat t$ can be determined even in the absence of calibration of the experiment.

The measurement of the third moment of current fluctuations provides information on the distribution of the transparencies of the channels of the sample. In particular, since it involves the third power of the $T_n$, it is much more sensitive to open channels than the conductance or even than $S_{I^2}$. For example, in tunnel junctions, the distribution  may have a tail that extends close to $t=1$ (pinholes) \cite{bardou}. Let us consider a tunnel junction of $M=10^8$ channels of transparency $t=10^{-5}$  (i.e., with a resistance about 26$\Omega$). This corresponds to ${\cal T}_p=t^p$. If one adds to the junction 100 channels of transparency $t'=0.1$, the distribution of the channels is modified. The average transmission ${\cal T}_1$, and thus the conductance, is increased by only 1\%. The second moment is multiplied by 10, but this modifies $\eta$ by only $10^{-3}$. In contrast, ${\cal T}_3$ is multiplied by $10^6$, so $\hat t\sim10^{-3}$, i.e., is multiplied by 100 ! For a tunnel junction, the temperature dependent part (the first term of Eq. (\ref{eq_S3bis}) is proportionnal to $\hat t$ and thus enhanced by a factor 100, but the second term, proportionnal to $(1-2\hat t)$ is not much influenced by the additionnal open channels.

As we show below, our samples correspond to good junction tunnel for which $t\sim10^{-5}$. Using the general property $\hat t\leq(\eta^{-1}-1)$, we deduce from our measurement of $\eta\geq0.95$, that $\hat t\leq0.05$. Thus, one can consider that our samples are tunnel junctions with an opaque barrier. For such a sample, the current noise spectral density (related below to the second moment) is given by : $S_{I^2}=eGV \rm{coth}(eV/2k_BT)$ . Only at high voltage $eV\gg k_BT$ does this reduce to the Poisson result $S_{I^2}=eI$. The third moment of the fluctuations is given by  $S_{I^3}=e^2I$.


\subsection{THE NOISE IN FOURIER SPACE}

The quantities $S_{I^p}$ are spectral densities, expressed in A$^p$/Hz$^{p-1}$. Until now we have supposed that $S_{I^p}$ is frequency independent, but in general it may depend on frequency. Then the $p$th moment depends on $p-1$ frequencies $f_1\dots f_{p-1}$. However, it is convenient to express $S_{I^p}$ as a function of $p$ frequencies such that the sum of all the frequencies is zero. Introducing the Fourier components $i(f)$ of the current, one has, for a classical current:
\begin{equation}
S_{I^p}(f_1,\dots,f_{p-1})=\left<i(f_1)\dots i(f_p)\right>\delta(f_1+\dots f_p)
\end{equation}
In quantum mechanics, the current operators taken at different times do not commute; they also do not in Fourier space, and the question of how the operators have to be ordered is crucial \cite{LevitovMath,Chtche}.

In the case of the second moment, one has $S_{I^2}(f)=\left<i(f)i(-f)\right>=\left<|i(f)|^2\right>$. It measures the power emitted by the sample at the frequency $f$ within a bandwidth of 1Hz. This is what a spectrum analyzer measures. Experimentally, the current emitted by the sample runs through a series of cables, filters and amplifiers before beeing detected (this can be avoided by an on-chip detection \cite{richard}). Thus, the measured quantity is a filtered current $j(f)=i(f)g(f)$ where $g(f)$ describes the filter function of the detection. 

One is often interested in the total power emitted by the sample. This is obtained by measuring the DC voltage after squaring $j(t)$, i.e. $\int j^2(t)dt$. This quantity is related to $S_{I^2}$ through:
\begin{equation}
\left<\delta I^2\right>=\int_{-\infty}^{+\infty}j^2(t)dt=
\int\int_{-\infty}^{+\infty}g(f_1)g(f_2)\left<i(f_1)i(f_2)\right>\delta(f_1+f_2)
=\int |g(f)|^2S_{I^2}(f)df
\end{equation}
It is remarkable that the frequency-dependent phase shift introduced by $g(f)$ has no influence, in agreement with the fact that $S_{I^2}$ has the meaning of a power.
If the detection bandwidth extend from $F_1$ to $F_2$, i.e., $g(f)=1$ for $F_1<|f|<F_2$ and $g(f)=0$ otherwise, and if $S_{I^2}$ is frequency independent between $F_1$ and $F_2$, then the total noise is given by $\left< \delta I^2\right>=2S_{I^2}(F_2-F_1)$. This corresponds to the number of choices of $f$ between $F_1$ and $F_2$, i.e., $F_2-F_1$, multiplied by the number of choices for the sign of $f$, i.e., 2.

Let us consider now $S_{I^3}$. As we said before, the spectral density of the third moment of current fluctuations depend on two frequencies $f_1$ and $f_2$, and the third moment of the measured current $j(t)$ is given by:
\begin{equation}
\left<\delta I^3\right>=\int j^3(t)dt=
\int\int\int g(f_1)g(f_2)g(f_3) S_{I^3}(f_1,f_2)\delta(f_1+f_2+f_3) df_1df_2df_3 \end{equation}
We see that now the phase of $g(f)$ matters. More precisely, $S_{I^3}$ measures how three Fourier components of the current can beat together to give a non-zero result, i.e., it measures the phase correlations between these three Fourier components. With the same hypothesis as before, for a detection between $F_1$ and $F_2$, one has now $\left<\delta I^3\right>=3S_{I^3}(F_2-2F_1)^2$ if $F_2>2F_1$ and $\left<\delta I^3\right>=0$ otherwise. This shows how important it is to make a broadband measurement. This unusual dependence of the result on $F_1$ and $F_2$ comes from the fact that the lowest frequency being the sum of two others is $2F_1$ whereas the maximum frequency one can subtract to that in order to have a DC signal is $F_2$. The number of choices for $f_1$ and $f_2$ is $(F_2-F_1)^2/2$ and the number of possible signs for the frequencies is 6 (all combinations except $+++$ and $---$).
 As we show below, we have experimentally confirmed this unusual dependence of $\left<\delta I^3\right>$ on $F_1$ and $F_2$. 

\subsection{EFFECT OF THE ENVIRONMENT}

\begin{figure}
\includegraphics[width= 0.5\columnwidth]{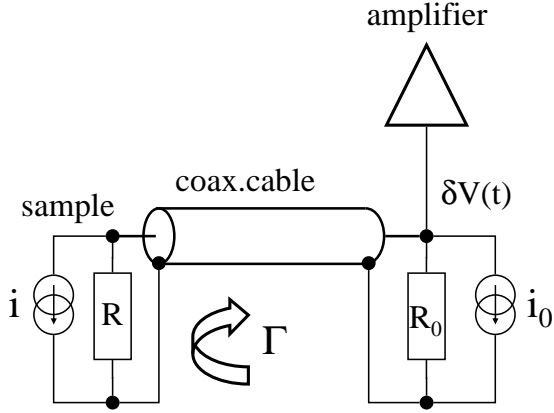}
\caption{Schematics of the equivalent circuit used for the theoretical model.}
\label{schematics}
\end{figure}

We now consider the effects of the sample's electromagnetic environment (contacts, leads, amplifier, etc.); the sample is no longer voltage biased. The environment emits noise, inducing fluctuations of the voltage across the sample, which in turn modify the probability distribution $P$. Moreover, due to the finite impedance of the environment, the noise emitted by the sample itself induces also voltage fluctuations. This self-mixing of the noise by the sample is responsible for the Coulomb blockade of high resistance samples, where it induces a modification of the $I(V)$ characteristics \cite{Ingold}. We consider the circuit depicted in Fig. \ref{schematics}, at first neglecting time delay along the coaxial cable. The noise of the sample of resistance $R$ is modeled by a current generator $i$. The voltage $\delta V$ is measured across a resistor $R_0$, which has a current generator $i_0$ of noise spectral density $S_{i_0^2}$. One has $\delta V=-R_D(i+i_0)$ with $R_D=RR_0/(R+R_0)$ ($R$ in parallel with $R_0$). The spectral density of the second moment of the voltage fluctuations is $S_{V^2}=+R_D^2(S_{I^2}+S_{i_0^2})$. Thus, the only effect of the environment on the second moment of the noise is to rescale the fluctuations (by $R_D^2$) and add a bias-independent contribution \cite{BuBlan}.
In contrast, it has been recently predicted that the third moment of $P$ is significantly modified by the environment  \cite{Kindermann2}:
\begin{equation}
S_{V^3}=-R_D^3S_{I^3}+3R_D^4S_{i_0^2}\frac{dS_{I^2}}{dV}+3R_D^4S_{I^2}\frac{dS_{I^2}}{dV}
\label{eqenv}
\end{equation}

We now give a simple derivation of this result. One has 
$\left<\delta V^3\right>=-R_D^3(\left<i^3\right>+3\left<i^2i_0\right>+3\left<ii_0^2\right>+\left<i_0^3\right>)$. The term $\left<ii_0^2\right>$ is zero. The term $\left<i_0^3\right>$ is zero for a environment at equilibrium, but can be non-zero if the noise that the sample sees comes from an active device like an amplifier. Our detection method is insensitive to this term and we will neglect it.
The first term on the right of Eq. (\ref{eqenv}) is like that of the second moment (it comes from $\left<i^3\right>$). The negative sign results from an increasing sample current giving a reduced voltage.  In order to understand the other terms, let us first consider the correlator $\left<i^2i_0\right>$. This correlator measures the effect of the external current $i_0$ on the amount of noise $i^2$ emitted by the sample. If $i_0$ varies slowly on the time scale of the current fluctuations (i.e., $\sim h/(eV)$), one can replace $i^2$ by its time average, $S_{I^2}$ taken at the instantaneous, slowly varying (with a time scale of the order of the RC time of the junction), voltage $V(t)=V+\delta V(t)$. Since $\delta V$ depends on $i_0$, the correlator $\left<i^2i_0\right>$ does not vanish. We can calculate it for $\delta V\ll k_BT$, for which one can Taylor expand $S_{I^2}$ in $S_{I^2}(V+\delta V(t))\simeq S_{I^2}(V)+\delta V(t)dS_{I^2}/dV$. We deduce $\left<i^2i_0\right>\simeq R_D\left<i_0^2\right>dS_{I^2}/dV$. This leads to the second term on the right of Eq.(\ref{eqenv}). It represents the effect of the noise emitted by the environment, and disappeares if the environment is at zero temperature. Similarly, the correlator $\left<i^3\right>$ can be seen as the effect of the sample's current $i$ acting on the noise $i^2$ emitted by itself, and leads to the third term on the right of Eq. (\ref{eqenv}) (the time scale are no longer separated, but a rigorous calculation \cite{Kindermann2} shows that Eq.(\ref{eqenv}) is nevertheless valid). This happens because of the finite impedance of the environment and persists at zero temperature. It represents the feedback of the environment. 
We give below a simple derivation of how to include the effect of propagation time in the coaxial cable, which dramatically affects $S_{V^3}$. 

Let us remark that there are terms of order $\delta V^2$ and higher, that are small and neglected here. Such terms affect slightly $S_{V^2}$ as well. They correspond to photon assisted noise in the low frequency limit $\hbar\omega\ll k_BT$ \cite{BuBlan,Rob}.

\vspace{3mm}
\section{THE EXPERIMENT}
\vspace{2mm}

\begin{figure}
\includegraphics[width= 0.8\columnwidth]{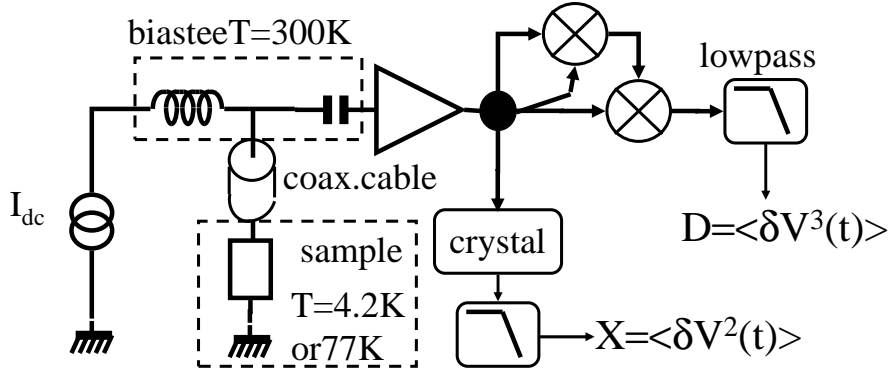}
\caption{Schematics of the experimental setup.}
\label{setup}
\end{figure}

\subsection{SIGNAL TO NOISE RATIO}

Let us evaluate the signal-to-noise ratio (SNR) for the measurement of $\left<\delta
I^3\right>$ as compared to the measurements of $\left<I\right>$ and $\left<\delta
I^2\right>$. We will suppose that the noise is dominated by the amplifier's noise current $I_A$, characterized by a spectral density $S_A$. We note $B$ the bandwidth of the measurement and $\tau$ the averaging time. We suppose the low frequency cutoff to be much smaller than the upper one, such that $\left<\delta I^p\right>\propto B^{p-1}$. The number of statistically independent measurements is $N=B\tau$, and the SNR always increases like $N^{1/2}$. For the dc current the SNR is :
\begin{equation}
\sigma_1=\frac{IN^{1/2}}{({\rm var} I_A)^{1/2}}=\frac{IN^{1/2}}{(2S_AB)^{1/2}}
=I\sqrt{\frac{\tau}{2S_A}}
\end{equation}
where var denotes the statistical variance, var$(x)=\left<x^2\right>-\left<x\right>^2$. We suppose the amplifier's noise to be gaussian, so var$(I_A^p)=a_p(\left<I_A^2\right>)^p=a_p(2S_AB)^p$.  $a_1=1$, $a_2=2$, $a_3=15$, etc. \footnote{The rapid decay of the SNR with $p$ could be avoided by the use of $p$ amplifiers and correlation methods.}
As expected, the result for $\sigma_1$ is independant of $B$, since $I$ is a dc quantity.
Similarly, one has for the second moment:
\begin{equation}
\sigma_2=\frac{(2eIB)N^{1/2}}{({\rm var} I_A^2)^{1/2}}=\sigma_1 x
\end{equation}
with $x^2=e(eB)/S_A$. $x^2$ is the ratio of the shot noise of a current eB (one electron every $B^{-1}$ seconds) to the noise of the amplifier.
Finally, the SNR $\sigma_3$ of the third moment is:
\begin{equation}
\sigma_3=\frac{(3e^2IB^2)N^{1/2}}{({\rm var} I_A^3)^{1/2}}=\frac3{2\sqrt{15}}\sigma_1 x^2
\end{equation}

For a sample with high impedance, one should use detection in the audio frequency. Taking $\tau=1$s, $B=10$kHz, $S_A=10^{-28}$A$^2$/Hz, gives $x\sim10^{-3}$. This estimate shows how difficult is the measurement of $S_{I^3}$ as compared to a conductance, or even a $S_{I^2}$ measurement. For a sample with an impedance comparable with $50\Omega$, one can use RF technique and have a huge bandwidth. We chose this option. Our setup corresponds to $B=1$GHz and a noise temperature of $\sim250$K ($S_A\sim3.\;10^{-22}$), and thus $x\sim3.\;10^{-4}$. This seems much worse than the previous case; however, one can pass much more current through a low resistance sample, which makes the two techniques comparable. The use of a cryogenic amplifier, with 10 times smaller noise temperature would increase the SNR by $10^{3/2}\sim30$. However, one can be limited by the amount of current through the sample, in particular if the relevant voltage scale is set by the temperature.

\subsection{THE SAMPLES}

Two samples have been studied. Both are tunnel junctions made of Al/Al oxide/Al, using the double angle evaporation technique \cite{FultonDolan}. In sample A the bottom and top Al films are 50 nm thick. The bottom electrode was oxidized for 2 hours in pure $O_2$ at a pressure of 500 mTorr. The junction area is $15\;\mu$m$^2$. In sample B, the films are 120 nm and 300 nm thick, oxidation was for 10 min, and the junction area is $5.6\mu m^2$ \cite{Lafe}.

\subsection{EXPERIMENTAL SETUP}

We have measured $\delta V(t)^3$ in real time (see Fig. \ref{setup}).
The third moment of the voltage fluctuations $S_{V^3}$ is a very small quantity, and its measurement requires much more care than $S_{V^2}$. For example, any non linearity of the amplifiers gives rise to a fake signal, and we have checked that our result was independent of the amplitude of the signal at the input of the various amplifiers.
Moreover, since the signal is very small, much more averaging is needed than for $S_{V^2}$.

\begin{figure}
\includegraphics[width= 0.8\columnwidth]{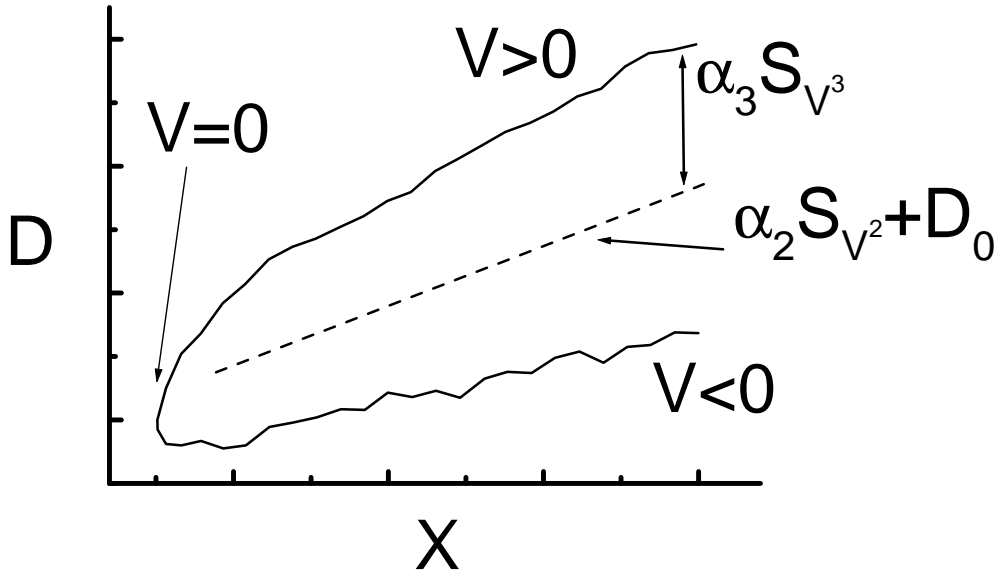}
\caption{D vs. X as the voltage across the sample is varied. For opposite voltage, X gives the same result ($X(-V)=X(V)$) whereas D doesn't. This proves the existence of the asymmetry of the fluctuations.}
\label{DofX}
\end{figure}

The sample is dc current biased through a bias tee. The noise emitted by the sample is coupled out to an rf amplifier through a capacitor so only the ac part of the current is amplified. The resistance of the sample is close to $50\;\Omega$, and thus is well matched to the coaxial cable and amplifier. After amplification at room temperature the signal is separated into four equal branches, each of which carries a signal proportionnal to $\delta V(t)$. A mixer multiplies two of the branches, giving $\delta V^2(t)$; a second mixer multiplies this result with another branch. The length of the cables between the splitter and the mixers has been adjusted to insure that the products correspond to the voltage taken at the same time. This has been done with the use of short (1ns long) pulses. The output of the second mixer, $\delta V^3(t)$, is then low pass filtered, to give a signal which we refer to as D. Ideally D is simply proportional to $S_{V^3}$, where the constant of proportionality depends on mixer gains and frequency bandwidth. The last branch is connected to a square-law crystal detector, which produces a voltage $X$ proportional to the the rf power it receives: the noise of the sample $\left<\delta V^2\right>$ plus the noise of the amplifiers. The dc current $I$ through the sample is swept slowly. We record $D(I)$ and $X(I)$; these are averaged numerically. This detection scheme has the advantage of the large bandwidth it provides ($\sim1$ GHz), which is crucial for the measurement. Due to the imperfections of the mixers, $D$ contains some contribution of $S_{V^2}$, and some contribution of the amplifiers, which are independent of $I$: $D=D_0+\alpha_3S_{V^3}+\alpha_2S_{V^2}$. When the current is reversed, one observes that $X$ is the same, $X(-I)=X(I)$, while $D$ is not, see Fig. \ref{DofX}. This prooves the asymmetry of the current fluctuations. We deduce $S_{V^3}=(D(I)-D(-I))/(2\alpha_3)$, since $S_{V^2}$ does not depend on the sign of $I$ \footnote{In the case this is not correct, one has to subtract any parasitic contribution, which is possible since the calibration procedure gives access to $\alpha_2$ and $\alpha_3$.}.

\subsection{CALIBRATION}

In order to calibrate the measurement of $\left<\delta V^2\right>$, we measure the Johnson noise emitted by a macroscopic $50\Omega$ resistor at various temperatures. This gives the gain of the detecion scheme as well as its noise temperature (dominated by the first amplifier). Since the third moment of the current fluctuations had never been measured before, we cannot rely on any sample. Thus,
in order to determine the magnitude and sign of $\left<\delta V^3\right>$ we measured the signal D when the sample is replaced  by a programmable function generator. The output of the generator consist of a pseudo-random sequence of voltage steps (at a rate of $10^9$ samples per second, 1GS/s), the statistics of which we calculated to be asymmetric in a known way. By varying the maximum amplitude $A$ of the signal and reversing its sign, we can measure the coefficients $\alpha_2$ and $\alpha_3$, which vary as $A^2$ and $A^3$ respectively. However, the actual signal synthesized by the generator is not exactly the one which one specifies. Due to the finite output bandwidth of the generator, the steps are rounded, which modifies the statistics of the output signal. We measured this statistics with a 20GS/s oscilloscope, and used the result to calibrate our measurement.

Now that we have checked what $S_{V^3}$ is for our sample, we can use the result to calibrate future experiments. In particular we can deduce from the measurement of a tunnel junction what are the parameters characterizing the environment (see later), which is essential if we want to study more exotic samples for which the intrinsic $S_{I^3}$ is unknown.

\vspace{3mm}
\section{EXPERIMENTAL RESULTS}
\vspace{2mm}

\subsection{DIFFERENTIAL RESISTANCE AND SECOND MOMENT OF NOISE}

Sample A was measured at $T=4.2$K. Its total resistance (tunnel junction and contacts) is $62.6\;\Omega$. The resistance of the junction itself is extracted from the fit of $S_{V^2}(V)$, which is a function of $eV/k_BT$ with $V$ the voltage drop across the junction. (The small resistance of the contacts adds only a voltage independent contribution to $S_{V^2}$). We find  $R_A=49.6\;\Omega$. We measured the differential resistance of the junction in the range of voltage that corresponds to the measurement of $S_{V^3}$. $R_{diff}$ is voltage independent to within  $1\%$. We extract the Fano factor $\eta$ of the sample from the slope of the $S_{V^2}(V)$ curve at high voltage. One gets $\eta=1$ (i.e., the Poisson result, corresponding to an opaque junction) with a precision of a few percent. Measurements on similar samples in the superconducting state show that the current at finite voltage below the gap $\Delta/e$ follows the BCS prediction \cite{Chris}, an indication that the junction has no open channels . 

\begin{figure}
\includegraphics[width= 0.7\columnwidth]{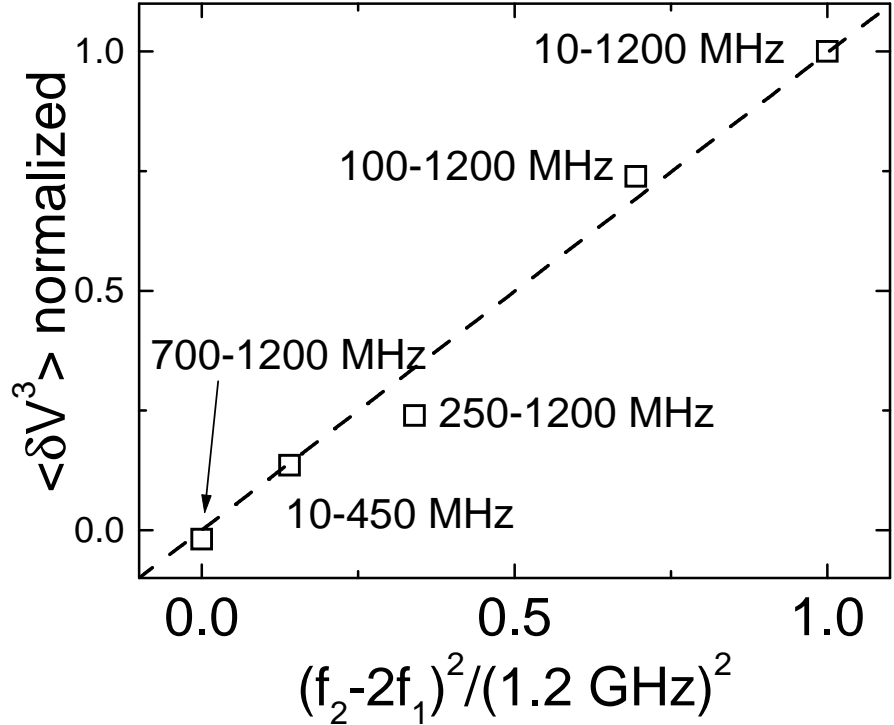}
\caption{Effect of finite bandwidth on the measurement of $\left<\delta V^3\right>$. Each point corresponds to a different value of the frequencies $F_1$ and $F_2$, as indicated in the plot. The data shown here correspond to sample B at $T=77$ K.}
\label{bandwidth}
\end{figure}

Sample B was measured at $T=4.2$ K, $T=77$ K and $290$ K. The contact resistance is $\sim1\Omega$ and the resistance of the junction is $86\;\Omega$, independent of temperature. $R_{diff}$ shows significant curvature on a voltage scale of 160 mV (the scale used at 300K and 77K): $R_{diff}$(-160mV)=83.5 $\Omega$, $R_{diff}$(+160mV)=82.5 $\Omega$. It is almost flat on the 16mV scale (the scale used at 4 K). This phenomenon is usual in tunnel junction and associated with the finite height and asymmetry of the barrier \cite{barrier}. The Fano factor of this sample is also $\eta=1$. Measurements of shot noise on similar samples in a different experimental setup lead to the same conclusion \cite{Lafe}.

\subsection{EFFECT OF THE BANDWIDTH ON THE THIRD MOMENT}

As we demonstrated before $S_{I^3}$ has an unusual dependence on the detection bandwidth, from $F_1$ to $F_2$. This property is also valid for $S_{V^3}$, even in the presence of an environment with frequency independent impedance. A general expression for that can be found in Ref.~\citenum{BigKindermann}.

A powerful check that $D$ really measures $S_{V^3}$ is given by varying the  bandwidth. The scaling of $S_{V^3}$ with $F_1$ and $F_2$ ($S_{V^3}\propto(F_2-2F_1)^2$ if $F_2>2F_1$ and 0 otherwise) is characteristic of the measurement of a third order moment. We do not know any experimental artifact that has such behavior.
 $F_1$ and $F_2$ are varied by inserting filters before the splitter. As can be seen in Fig. \ref{bandwidth}, our measurement follows the dependence on $(F_2-2F_1)^2$, which cannot be cast into a function of $(F_2-F_1)$.
Each point on the curve of Fig. \ref{bandwidth} corresponds to a full $\left<\delta V^3(I)\right>$ measurement (see figures \ref{samplea} and \ref{sampleb}). An asymmetric external noise would just add just a constant to $D$, and thus would not affect our measurement, even though this constant might have the same dependence on bandwidth as $\left<\delta V^3\right>$.

\subsection{RESULT FOR THE THIRD MOMENT}

\begin{figure}
\includegraphics[width= 0.8\columnwidth]{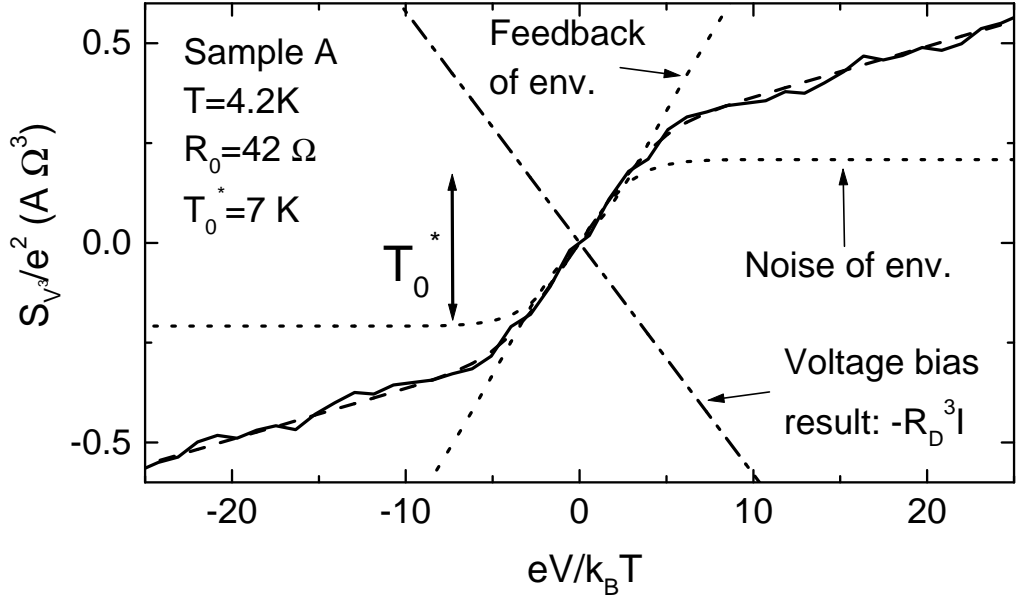}
\caption{Measurement of $S_{V^3}(eV/k_BT)$ for sample A (solid line). The dashed line corresponds to the best fit with Eq. (\ref{eqenv}). The dash dotted line corresponds to the perfect bias voltage contribution and the dotted lines to the effect of the environment.}
\label{samplea}
\end{figure}

\begin{figure}
\includegraphics[width= 0.7\columnwidth]{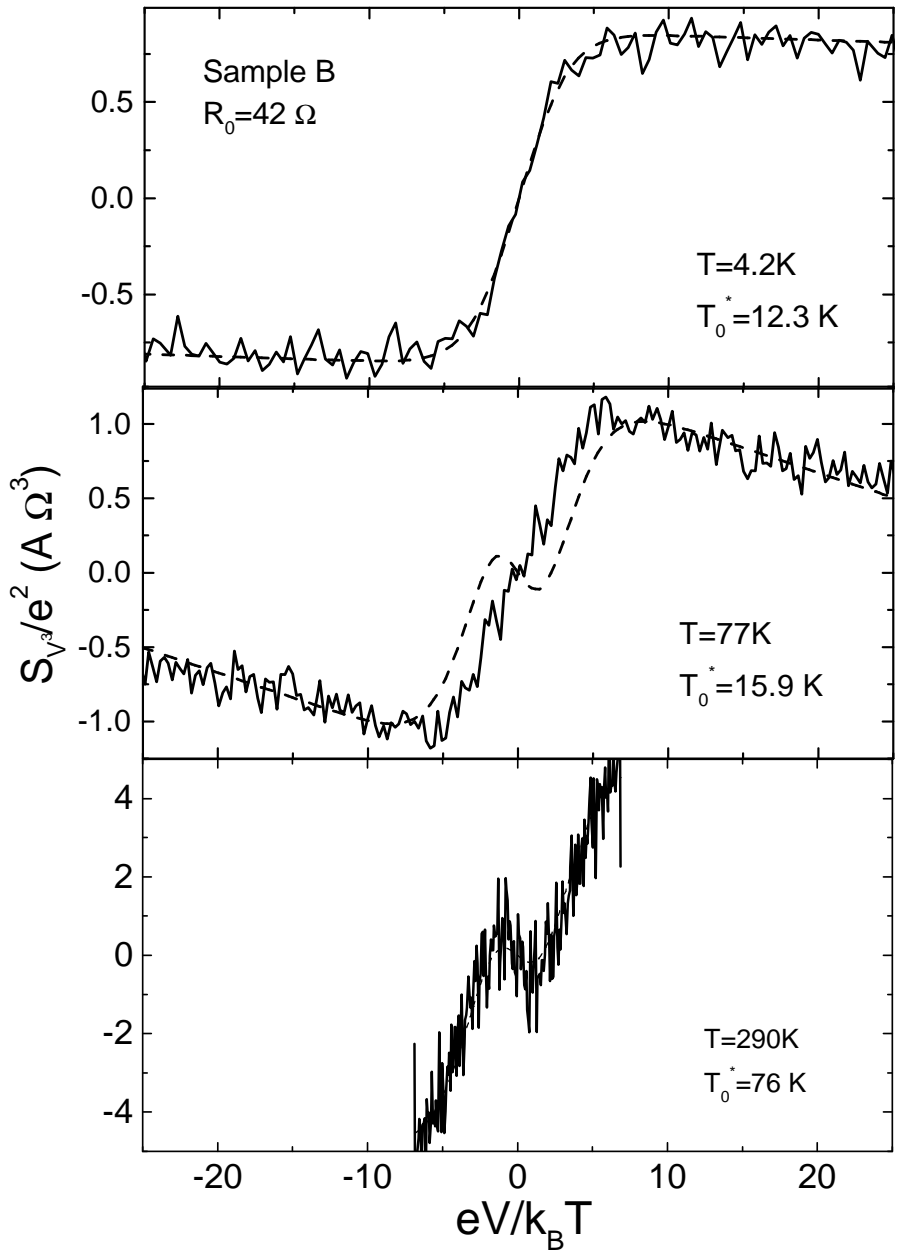}
\caption{Measurement of $S_{V^3}(eV/k_BT)$ for sample B (solid lines). The dashed lines corresponds to the best fit with Eq. (\ref{eqenv}).}
\label{sampleb}
\end{figure}

$S_{V^3}(eV/k_BT)$ for sample A at $T=4.2$K is shown in Fig. \ref{samplea}; these data were averaged for 12 days.
$S_{V^3}(eV/k_BT)$ for sample B at $T=4.2$K (top), $T=77$K (middle) and $T=290$K (bottom) is shown in Fig. \ref{sampleb}. The averaging time for each trace was 16 hours.
These results are clearly different from the voltage bias result (the dash-dotted line in Fig. \ref{samplea}). However, all our data are very well fitted by Eq. (\ref{eqenv}) which takes into account the effect of the environment (see the dash lines of Fig. \ref{samplea} and \ref{sampleb}).
The unknown parameters are the resistance $R_0$ and the environment effective noise temperature $T_0^*$ (as we will see the latter does not correspond to the real noise temperature of the environment). We checked that the impedance of the samples was frequency independent up to 1.2 GHz within  5\%.  Fig. \ref{samplea} and \ref{sampleb} show the best fits to the theory, Eq. (\ref{eqenv}), for all our data. The four curves lead to $R_0=42\;\Omega$, in agreement with the fact that 
the electromagnetic environment (amplifier, bias tee, coaxial cable, sample holder) was identical for the two samples.
We have measured the impedance $Z_{env}$ seen by the sample. Due to impedance mismatch between the amplifier and the cable, there are standing waves along the cable. This causes $Z_{env}$ to be complex with a phase that varies with frequency. We measured that the modulus $|Z_{env}|$ varies between $30\;\Omega$ and $70\;\Omega$ within the detection bandwidth, in reasonable agreement with $R_0=42\;\Omega$ extracted from the fits. $R_0$ represents an average over frequency of the impedance seen by the sample\cite{BigKindermann}. We notice that the theory predicts a structure at low voltage for high enough temperature (see Fig. \ref{sampleb}, dashed line, middle and bottom), that we observe only at room temperature. We think this comes from the fact that at 77K we used a long cable, and thus had a largely frequency dependent impedance $Z_{env}$, whereas at room temperature we used a much shorter cable. The presence of the low voltage structure is indeed very much dependent on the value chosen for $R_0$, and might be washed out by frequency averaging with a long cable.

We have measured directly the noise emitted by the room temperature amplifier; we find $T_0\sim100$ K. This is in disagreement with the parameters $T_0^*$ deduced from the fit of the data. However, we show below that this is explained by the finite propagation time along the coaxial cable between the sample and the amplifier.

\subsection{EFFECT ON $T_0^*$ OF THE CABLE PROPAGATION TIME}

To analyze our results, let us consider again the circuit in Fig. \ref{schematics}, a simplified equivalent of our setup. $R_0\sim50\;\Omega$ is the input impedance of the amplifier, which is connected to the sample through a coaxial cable of impedance $R_0$ (i.e., matched to the amplifier) . The sample's voltage reflection coefficient is $\Gamma=(R-R_0)/(R+R_0)$. In the analysis we present next we neglect the influence of the contact resistance and impedance mismatch of the amplifier, but we have included it when computing the theory to compare to the data. The voltage $\delta V(t)$ measured by the amplifier at time $t$  arises from three contributions: i) the noise emitted by the amplifier at time $t$: $R_0i_0(t)/2$ ; half of $i_0$ enters the cable. ii) the noise emitted by the sample (at time $t-\Delta t$, where $\Delta t$ is the propagation delay along the cable) that couples into the cable: $(1-\Gamma)Ri(t-\Delta t)/2$ ; iii) the noise emitted by the amplifier at time $t-2\Delta t$ that is reflected by the sample: $\Gamma R_0i_0(t-2\Delta t)/2$ ; thus,
\begin{equation}
\delta V(t)=-\frac{R_0}2\left[i_0(t)+\Gamma i_0(t-2\Delta t)\right]-\frac R2(1-\Gamma)i(t-\Delta t)
\label{eqgamma}
\end{equation}
For $\Delta t=0$, Eq. (\ref{eqgamma}) reduces to $\delta V=-R_D(i+i_0)$ with $R_D=RR_0/(R+R_0)$. Thus, Eq.(\ref{eqgamma}) for $\Delta t=0$ corresponds to Eq. (\ref{eqenv}), which is a particular case of Eq. (12b) of Ref.~\citenum{Kindermann2}.
. Thus, $\left<\delta V^3\right>=-R_D^3(\left<i^3\right>+3\left<i^2i_0\right>+3\left<ii_0^2\right>+\left<i_0^3\right>)$ for $\Delta t=0$. As we already show, the term $\left<i^2i_0\right>$ leads to the second term on the right of Eq. (\ref{eqenv}). 
and the term $\left<i^3\right>$ to the first term of Eq. (\ref{eqenv}), and, due to the sample noise modulating its own voltage, the third term of Eq. (\ref{eqenv}) as well.

The finite propagation time does affect the correlator $\left<i^2i_0\right>$, which has to be replaced by $\left<i^2(t-\Delta t)i_0(t)\right>$.
Thus the term $S_{i_0^2}$ in Eq. (\ref{eqenv}) has to be replaced by $(\Gamma S_{i_0^2}+S_{i_0(t)i_0(t-2\Delta t)})/(1+\Gamma)$, where $S_{i_0(t)i_0(t-2\Delta t)}$ is the spectral density corresponding to the correlator $\left<i_0(t)i_0(t-2\Delta t)\right>$. For long enough $\Delta t$ this term vanishes, since $i_0$ emitted at times $t$ and $t-2\Delta t$ are uncorrelated. Thus, the effect of the propagation time is to renormalize the noise temperature of the environment $T_0=R_0S_{i_0^2}/(2k_B)$ into $T_0^*=T_0\Gamma/(1+\Gamma)$.

We now check whether the finite propagation time can explain our values of $T_0^*$.  The cable of length $\sim2$m corresponds to $\Delta t$ being long for the bandwidth we used. As a consequence, the relevant noise temperature to be used to explain the data is $T_0^*$. For sample A, $\Gamma=0.11$; including the contact resistance and cable attenuation one expects  $T_0^*=5$ K ; for sample B, $\Gamma=0.26$ and one expects $T_0^*=21$ K. A much shorter cable was used for $T=290$ K, and the reduction of $T_0$ is not complete. These numbers are in reasonably good agreement with the values of $T_0^*$ deduced from the fits (see Fig. \ref{samplea} and \ref{sampleb}), and certainly agree with the trend seen for the two samples. Clearly $T_0^*\ll T_0$ for the long cable.

\begin{figure}
\includegraphics[width= 0.8\columnwidth]{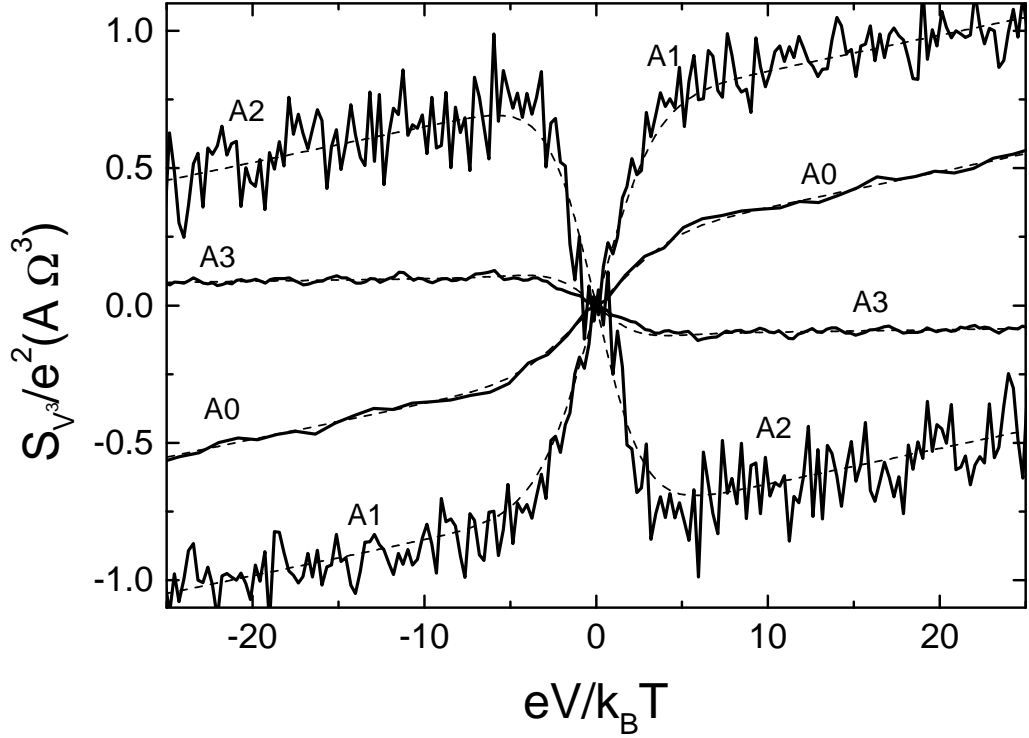}
\caption{Measurement of $S_{V^3}(eV/k_BT)$ for sample A at T=4.2 K (solid lines). A0: no ac excitation (same as Fig. \ref{samplea}). A1: with an ac excitation at frequency $\Omega/2\pi$ such that $\cos2\Omega\Delta t=+1$; A2: $\cos2\Omega\Delta t=-1$; A3: no ac excitation but a $63\;\Omega$ resistor in parallel with the sample. The dashed lines corresponds to fits with Eq. (\ref{eqenv}). }
\label{env}
\end{figure}

If there are some open channels in the tunnel junction, $S_{I^3}$ is modified according to Eq. (\ref{eq_S3bis}). Since we know that $\hat t$ is small, the effect on the temperature independent part of $S_{V^3}$ is small. However, the term proportionnal to $\hat t$ might influence $S_{V^3}$. It appears that this term is indistinguishable from the effect of the noise of the environnement (the correlator $\left<i^2i_0\right>$). It adds a (negative) contribution to $T_0$, given by $\delta T_0=-\hat tT(1+R_0/R)$. Taking $\hat t=0.05$ (the maximum value compatible with our measurement of $\eta$), one has $\delta T_0\sim-0.08T$. This effect is intrinsic to the sample and is not modified by the propagation delay. As a result, it would give an even better agreement between the evaluation of $T_0^*$ and the values deduced from the fit. A precise measurement of $\hat t$ would require a quieter environment (i.e., the use of a cryogenic amplifier) and/or a better impedance match between the sample and the amplifier.

\vspace{3mm}
\section{MODIFICATION OF THE ENVIRONMENT}
\vspace{2mm}

In order to demonstrate more explicitly the influence of the environment on $S_{V^3}$ we have modified the parameters $T_0$ and $R_0$ of the environment and measured the effect on $S_{V^3}$. 

$T_0$ is a measure of the current fluctuations emitted by the environment towards the sample. Its influence on $S_{V^3}$ is through the correlator $\left<i^2i_0\right>$. This correlator does not require $i_0$ to be a randomly fluctuating quantity in order not to vanish. So we can modify it by adding a signal $A\sin \Omega t$ to $i_0$ (with $\Omega$ within the detection bandwidth). That way we have been able to modify $T_0^*$ without changing $R_0$, as shown on Fig. \ref{env}. The term $S_{i_0(t)i_0(t-2\Delta t)}$ oscillates like $\cos2\Omega\Delta t$, and thus one can enhance (curve A1 as compared to A0 in Fig. \ref{env}) or decrease $T_0^*$, and even make it negative (Fig. \ref{env}, A2). The curves A0--A2 are all parallel at high voltage, as expected, since the impedance of the environment remains unchanged; $R_0=42\;\Omega$ is the same for the fit of the three curves. The experiment described here, i.e. measuring the current fluctuations in the presence of a dc plus an ac voltage, is called photon assisted noise when $S_{V^2}$ is measured\cite{Rob}. In this context, a frequency of the order of $eV/h$ has to be used, and the mechanism can be understood in terms of photons. In our case, since the third moment of the current fluctuations involves an odd number of current operators, the particle-like aspect of the ac excitation is less clear and deserves more study.

Second, by adding a $63\;\Omega$ resistor in parallel with the sample (without the ac excitation) we have been able to change the resistance of the environment $R_0$, and thus the high voltage slope of $S_{V^3}$. The fit of curve A3 gives $R_0=24.8\;\Omega$, in excellent agreement with the expected value of $25.2\;\Omega$ ($63\;\Omega$ in parallel with $42\;\Omega$). Since this makes the reflection coefficient negative ($\Gamma=-0.22$), the presence of the extra resistor also reverses the sign of $T_0^*$, as expected. This is not due to a negative temperature, but rather to a negative $\Gamma$.

\vspace{3mm}
\section{CONCLUSION}
\vspace{2mm}

Our data are fully consistent with a third moment of current fluctuations $S_{I^3}$ being independent of $T$ between 4K and 300K when the sample is voltage biased, as predicted for a tunnel junction. The effect of the environment, through its noise and impedance, is clearly demonstrated. This is of prime importance for designing future measurements on samples with unknown third moment.

\vspace{3mm}
\section*{ACKNOWLEDGMENTS}
\vspace{2mm}
We thank C. Beenakker, W. Belzig, A. Clerck, M.~Devoret, Y. Gefen, S. Girvin, N. Hengartner, M. Kindermann, L. Levitov, A. Mukherjee, Y. Nazarov and P.-E. Roche for useful discussions. This work was supported by NSF DMR 0072022 and the W.M. Keck Foundation.

\end{document}